\documentclass[%
prd%
 ,twocolumn%
 ,secnumarabic%
,amssymb, amsmath,nobibnotes, aps, prl, superscriptaddress]{revtex4}
\usepackage{docs}%
\usepackage{bm}%
\usepackage{graphicx}
\usepackage{color}
\usepackage{subfigure}
\expandafter\ifx\csname package@font\endcsname\relax\else
 \expandafter\expandafter
 \expandafter\usepackage
 \expandafter\expandafter
 \expandafter{\csname package@font\endcsname}%
\fi

\begin{document}
\title{Reduction of coating thermal noise by using an etalon}
\author{Kentaro Somiya}
\affiliation{Waseda Institute for Advanced Study, 1-6-1 Nishiwaseda, Shinjuku, Tokyo 169-8050, Japan}
\author{Alexey G. Gurkovsky}
\affiliation{Faculty of Physics, Moscow State University, Moscow, 119991 Russia}\author{Daniel Heinert}
\affiliation{Institut f\"ur Festk\"orperphysik, Friedrich-Schiller-Universit\"at Jena, D-07743 Jena, Germany}
\author{Stefan Hild}
\affiliation{SUPA, School of Physics and Astronomy, Institute for Gravitational Research, Glasgow University, Glasgow G12 8QQ, United Kingdom}
\author{Ronny Nawrodt}
\affiliation{Institut f\"ur Festk\"orperphysik, Friedrich-Schiller-Universit\"at Jena, D-07743 Jena, Germany}
\author{Sergey P. Vyatchanin}
\affiliation{Faculty of Physics, Moscow State University, Moscow, 119991 Russia}
\begin{abstract}
Reduction of coating thermal noise is a key issue in precise measurements with an optical interferometer. A good example of such a measurement device is a gravitational-wave detector, where each mirror is coated by a few tens of quarter-wavelength dielectric layers to achieve high reflectivity while the thermal-noise level increases with the number of layers. One way to realize the reduction of coating thermal noise, recently proposed by Khalili, is the mechanical separation of the first few layers from the rest so that a major part of the fluctuations contributes only little to the phase shift of the reflected light. Using an etalon, a Fabry-Perot optical resonator of a monolithic cavity, with a few coating layers on the front and significantly more on the back surface is a way to realize such a system without too much complexity, and in this paper we perform a thermal-noise analysis of an etalon using the Fluctuation-dissipation theorem with probes on both sides of a finite-size cylindrical mirror.
\end{abstract}

\maketitle

\section{Overview}

Brownian fluctuation of multi-layer coatings on a mirror is a dominant noise source in interferometric gravitational-wave detectors. One way to reduce coating thermal noise is mechanical separation of the first few layers from the rest of the coatings~\cite{Khalili}\cite{SomiyaPRL}. This so-called Khalili-cavity will be useful in precision measurements like the sub-SQL measurement experiment~\cite{10m} or a future-generation gravitational-wave detector like Einstein Telescope~{\cite{ET}\cite{ET2}}. Replacing a single mirror by a cavity will, however, increase the complexity of the system, and a way to ease this problem is to use an etalon instead of the double-mirror cavity. The front surface and the back surface of the etalon are not mechanically separated, but as we will show in this paper, the separation is good enough to reduce the noise level of the whole system better than a single conventional mirror.

\begin{figure}[htbp]
    \begin{center}
        \includegraphics[width=6cm]{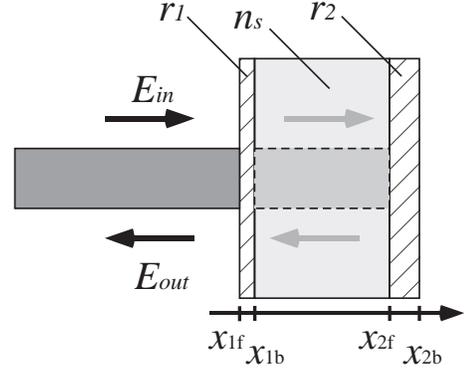}
    \caption{Khalili etalon with a few coating layers on the front surface and more layers on the back surface. The light is anti-resonant in the etalon so that a larger fraction of the light incident onto the etalon is reflected by the front surface. Mechanical losses of the back-surface coatings cause phase noise on the probe beam not only via thermal motion of the back surface, which is measured less, but also via thermal motion of the front surface as the two surfaces are connected by the substrate with finite thickness.}
    \label{fig:mirror}
    \end{center}
\end{figure}

Figure~\ref{fig:mirror} shows our model. The test mass is a cylindrical mirror made of silica and its motion is probed by an axisymmetric Gaussian beam. Both sides of the mirror are coated by silica-tantala doublets and the distance of the two surfaces is controlled to be anti-resonant for the carrier light utilizing the temperature dependence of the refractive index of the substrate. A larger fraction of the light is reflected back by the front surface but the thermal-noise level on the front surface is lower than that of the back surface because of the fewer coating layers. Each surface of the etalon is probed less than a fully reflective surface of a conventional mirror; the front surface of the etalon is probed by $|\epsilon_1|<1$, and the back surface is probed by $|\epsilon_2|<1$.

Total coating thermal noise of the etalon is the sum of noise on the front surface and noise on the back surface, which are caused by the thermal energy stored in the coating layers. In fact, it is not only the energy stored in the front (back) surface coatings that causes the phase fluctuation on the beam reflected at the front (back) surface of the etalon. The thermal energy stored in the coatings of the other surface contributes to the phase fluctuation on the beam via the substrate. For the etalon, we should thus consider four elements to calculate total coating thermal noise. One is the fluctuation of the front surface caused by the thermal energy in the front surface coatings. This process is the same as that of coating thermal noise of a conventional mirror. Second is the fluctuation of the front surface caused by the thermal energy in the back surface coatings that is {\it mechanically transferred} to the front surface. The other two are the fluctuation of the back surface coatings caused by the thermal energy in the front surface coatings and that in the back surface coatings, which are probed by the beam transmitting through the substrate or in other words {\it optically transferred} to the front surface. The fluctuations caused by the thermal energy in the coatings of a same surface are correlated.

According to the Fluctuation-dissipation theorem, the fluctuation caused by the thermal energy is given by calculating dissipation of the elastic energy caused by an imaginary force that is intentionally applied to the measured surface~{\cite{FdT}\cite{Levin}}. In the case of a conventional mirror, a single imaginary force is applied to the reflective front surface, and the elastic energy is integrated over the coatings on the surface. In the case of an etalon, an imaginary force shall be applied to each reflective plane with a weight coefficient $\epsilon_j$, which is given from the response from the motion of each surface to the phase change of the reflected light. The elastic energy is then integrated over the coatings on both surfaces. The thermal-noise level is proportional to the dissipated power, which is given by the elastic energy multiplied by the mechanical loss angle.

To calculate the elastic energy of the etalon, we should solve the elastic equation of the finite-size cylinder. As is done in Ref.~\cite{GEOBS}, we use Levin's method~\cite{Levin} with one probe beam on the front and another probe beam on the back surface. The way to extend Levin's method to a finite-size mirror has been shown by Bondu {\it et al.}~{\cite{Bondu}\cite{Liu}\cite{SomiyaTN}}, and we follow the same way with slightly different boundary conditions using the two probe beams; a similar approach has been introduced without the derivation in Ref.~\cite{Numata} to evaluate thermal noise of a monolithic cavity. We will show the semi-analytical derivation for the model with the two probe beams in Sec.~\ref{sec:thinLayer}. In fact, for this model, the front surface coatings are regarded to be so thin that the difference between the front side and the rear side of the coating layers is negligible -- {\it thin-layer model}. A closer look into the optical behavior reveals that the light transmitting through the front surface coatings circulates inside the etalon with reflecting on the back surface coatings and on the {\it rear side of the front surface coatings}. In Sec.~\ref{sec:thickLayer}, we take the difference of the two sides of the front surface coatings into account -- {\it thick-layer model}. Section~\ref{sec:results} shows the result of the calculation. We have verified the result by numerical calculation using a finite-element analysis code, which is shown in Sec.~\ref{sec:discussion} with other discussions.

\section{Thin-layer model}
\label{sec:thinLayer}

Figure~\ref{fig:mirror} shows our model. As is done in Ref.~{\cite{SomiyaTN}\cite{Harry}}, the multi-layer silica-tantala doublets are approximated as a single thick layer of silica or tantala. The dissipations are calculated for silica and tantala to be square-summed (see Sec.~\ref{sec:multi} for details). The reflectivity of the approximated mono-layer coating is that of the multi-layer coatings. 

In the thin-layer model, we assume that the light is always reflected at the front side of the coatings. This is the case for a conventional mirror where the light transmitting through the coatings never comes back. This is also true for the back surface coatings. The question is about the front surface coatings, but let us ignore the difference for the simplicity. This simplification is valid as the anti-resonant etalon accommodates little light circulating in it.

\subsection{Probe force amplitudes $\epsilon_j$}
We can use the quasi-static approximation for a short cavity, which means that displacements of mirrors are sufficiently slow compared with the relaxation rate of the cavity. The optical path length between the surfaces can be assumed to be an odd multiple of the quarter-wavelength in the zeroth order. Phase of the light reflected by the cavity as a compound mirror changes according to the differential motion of the two mirrors $\delta x_\mathrm{2f}-\delta x_\mathrm{1f}$. In addition, the phase shift due to the motion of the front mirror from its initial position should be taken into account. Note that the former motion is probed inside the substrate, thus the phase shift is proportional to the refractive index of the silica substrate $n_s$. The reflected light field $E_{out}$ can be given with the input field $E_{in}$ as
\begin{eqnarray}
\frac{E_{out}}{E_{in}}&=&-e^{2ik_0\delta x_\mathrm{1f}}\frac{r_1+r_2 e^{2i\theta}}{1+r_1r_2e^{{2i\theta}}}\ ,
\label{eq:BandA} \\
\theta&=&k_0n_s(\delta x_\mathrm{2f}-\delta x_\mathrm{1f})\ , \nonumber
\end{eqnarray}
$r_j$ is the amplitude reflectivity of the front/back surface, and $k_0$ is the wave number. Expanding Eq.~(\ref{eq:BandA}) into a series over $\delta x_\mathrm{1f}$ and $\delta x_\mathrm{2f}$, and keeping linear terms, we obtain
\begin{eqnarray}
\frac{E_{out}}{E_{in}}\simeq-\frac{r_1+r_2}{1+r_1r_2}-2ik_0\left( \epsilon_\mathrm{1}\delta x_\mathrm{1f} + \epsilon_\mathrm{2}\delta x_\mathrm{2f}\right)
\end{eqnarray}
with
\begin{eqnarray}
\epsilon_\mathrm{1}&=&\frac{r_1+r_2}{1+r_1r_2}-\frac{n_sr_2(1-r_1^2)}{(1+r_1r_2)^2}\ ,\nonumber\\
\epsilon_\mathrm{2}&=&\frac{n_sr_2(1-r_1^2)}{(1+r_1r_2)^2}\ .\label{eq:epsilon}
\end{eqnarray}
These $\epsilon_j$ are used as weight coefficients of the imaginary forces applied to the surfaces of the etalon.

\subsection{Elastic motion of the substrate}
\label{subsec:thinlayerelastic}

The Fluctuation-dissipation theorem tells us that the power spectrum of thermal noise at the radial frequency $\Omega$ is given by~\cite{Levin}
\begin{eqnarray}
S_x(\Omega)=\frac{8k_\mathrm{B}T}{\Omega F_0^2}U\phi\ ,\label{eq:FdT}
\end{eqnarray}
where $k_\mathrm{B}$ is the Boltzmann constant, $T$ is the temperature, $U$ is the maximum elastic energy that can be generated by the imaginary force $F_0$, and $\phi$ is the loss angle. 

While the conventional method applies a single imaginary force on the only reflective surface of a mirror, our new method shall apply imaginary forces on both sides of the etalon. The weighting coefficients of the imaginary force are $\epsilon_1$ and $\epsilon_2$ that we have derived above.

The elastic energy is given by the product of the strain tensor $E_{ij}$ and the stress tensor $T_{ij}$, integrated over the volume of interest:
\begin{eqnarray}
U=\frac{1}{2}\int\!\sum_{i,j}E_{ij}T_{ij}dV\ \ \ \ (i,j=r,\psi,z)\ .\label{eq:U}
\end{eqnarray}
We use a cylindrical coordinate system along the $z$ axis; $r$ is the distance from the $z$ axis and $\psi$ is the angle around the $z$ axis. Our mass is a cylinder with radius of $a$ and thickness of $h$. The strain tensor elements of the cylinder with the axisymmetric pressure are expressed by the displacement vectors $u_r$ and $u_z$ as follows:
\begin{eqnarray}
&&E_{rr}=\frac{\partial u_r}{\partial r}\ ,\ \
E_{\psi\psi}=\frac{u_r}{r}\ ,\ \
E_{zz}=\frac{\partial u_z}{\partial z}\ ,\nonumber\\
&&E_{rz}=\frac{1}{2}\left(\frac{\partial u_r}{\partial z}+\frac{\partial u_z}{\partial r}\right)\ ,\label{eq:strain}
\end{eqnarray}
and the stress tensor elements are as follows:
\begin{eqnarray}
T_{rr}&=&(\lambda+2\mu)E_{rr}+\lambda(E_{\psi\psi}+E_{zz})\ ,\nonumber\\
T_{\psi\psi}&=&(\lambda+2\mu)E_{\psi\psi}+\lambda(E_{zz}+E_{rr})\ ,\nonumber\\
T_{zz}&=&(\lambda+2\mu)E_{zz}+\lambda(E_{rr}+E_{\psi\psi})\ ,\nonumber\\
T_{rz}&=&2\mu E_{rz}\ .\label{eq:stress}
\end{eqnarray}
Here $\lambda$ and $\mu$ are the Lam\'e coefficients:
\begin{eqnarray}
\lambda=\frac{Y\nu}{(1+\nu)(1-2\nu)}\ ,\ \ \mu=\frac{Y}{2(1+\nu)}\ ,
\end{eqnarray}
with $Y$ the Young's modulus and $\nu$ the Poisson's ratio. According to Ref.~\cite{Bondu}, displacement vectors of a finite-size cylinder are described with the Bessel Functions:
\begin{eqnarray}
u_r(r,z)&=&\sum_m{A_m(z)}J_1(kr)+\delta u_r\label{eq:ur}\ ,\\
u_z(r,z)&=&\sum_m{B_m(z)}J_0(kr)+\delta u_z\ .\label{eq:uz}
\end{eqnarray}
$A_m$ and $B_m$ are in the same form as in Ref.~\cite{Bondu}:
\begin{eqnarray}
A_m(z)&=&\gamma_m e^{-kz}+\delta_m e^{kz}\nonumber\\
&&\hspace{1cm}+\frac{kz}{2}\frac{\lambda+\mu}{\lambda+2\mu}[\alpha_m e^{-kz}+\beta_m e^{kz}]\ ,\nonumber\\
\\
B_m(z)&=&\left(\frac{\lambda+3\mu}{2(\lambda+2\mu)}\alpha_m+\gamma_m\right)e^{-kz}\nonumber\\
&&\hspace{1cm}+\left(\frac{\lambda+3\mu}{2(\lambda+2\mu)}\beta_m-\delta_m\right)e^{kz}\nonumber\\
&&\hspace{1cm}+\frac{kz}{2}\frac{\lambda+\mu}{\lambda+2\mu}[\alpha_m e^{-kz}-\beta_m e^{kz}]\ .\nonumber\\
\end{eqnarray}
$\delta u_r$ and $\delta u_z$ are given as
\begin{eqnarray}
\delta u_r&=&a_1r+a_2rz\ ,\\
\delta u_z&=&b_1 z^2+b_2 r^2+b_3z\ .
\end{eqnarray}
The coefficients $\alpha_m$, $\beta_m$, $\gamma_m$, $\delta_m$, $a_1$, $a_2$, $b_1$, $b_2$, and $b_3$ are determined with the following boundary conditions:
\begin{align}
 T_{zz}(r,0)&=-\epsilon_1F_0p(r) \ , && \text{(i)}  \nonumber \\
 T_{zz}(r,h)&=\epsilon_2 F_0p(r) \ , && \text{(ii)} \nonumber \\
 T_{rz}(r,0)&=0                  \ ,    && \text{(iii)}  \nonumber \\
 T_{rz}(r,h)&=0                  \ ,    && \text{(iv)}  \nonumber \\
 T_{rz}(a,z)&=0                  \ , && \text{(v)} \nonumber \\
 T_{rr}(a,z)&=0                  \ . && \text{(vi)}
\end{align}
We have $\epsilon_j$ on the right side of the boundary conditions (i), (ii); the thermal-noise level of a single mirror will be given if we put $\epsilon_1=1$ and $\epsilon_2=0$. The beam profile $p(r)$ is expanded with Bessel functions:
\begin{eqnarray}
p(r)&=&\sum_m{p_mJ_0(k_mr)}+p_0\ ,\\
p_m&=&\frac{\exp{[-k_m^2w^2/8]}}{\pi a^2J_0^2(k_ma)}\ ,\\
p_0&=&\frac{1}{\pi a^2}\ .
\end{eqnarray}
Here $k_m\ (\rightarrow k)$ are the zeros of the first-order Bessel function, divided by the mirror radius $a$; namely $J_1(ka)=0$. The $p_0$ term is missed in Ref.~\cite{Bondu} and is corrected in Ref.~\cite{Liu}. Boundary condition (i) reads
\begin{eqnarray}
\left \{
\begin{array}{@{\,}c@{\,}}
k\mu[(\alpha_m-\beta_m)+2(\gamma_m+\delta_m)]=\epsilon_1F_0p_m\\
(\lambda+2\mu)b_3+2\lambda a_1=-\epsilon_1F_0p_0\ .
\end{array}\right .\label{eq:BC1}
\end{eqnarray}
Boundary condition (ii) reads
\begin{eqnarray}
\left \{
\begin{array}{@{\,}c@{\,}}
\displaystyle k\mu\bigg[(\alpha_m e^{-kh}-\beta_m e^{kh})+2(\gamma_m e^{-kh}+\delta_m e^{kh})\nonumber\\
\displaystyle \ \ \ +\frac{\lambda+\mu}{\lambda+2\mu}kh(\alpha_m e^{-kh}+\beta_m e^{kh})\bigg]=-\epsilon_2 F_0p_m\\
(\lambda+2\mu)(2b_1h+b_3)+2\lambda(a_1+a_2h)=\epsilon_2 F_0p_0\ .
\end{array}\right .\nonumber\\
\label{eq:BC2}
\end{eqnarray}
Boundary condition (iii) reads
\begin{eqnarray}
2k\mu\left[\frac{\mu}{\lambda+2\mu}(\alpha_m+\beta_m)+2(\gamma_m-\delta_m)\right]=0\ .\label{eq:BC3}
\end{eqnarray}
Boundary condition (iv) reads
\begin{eqnarray}
&&2k\mu\bigg[\frac{\mu}{\lambda+2\mu}(\alpha_m e^{-kh}+\beta_m e^{kh})\nonumber\\
&&\ \ \ \ \ \ \ \ \ +2(\gamma_m e^{-kh}-\delta_m e^{kh})\nonumber\\
&&\ \ \ \ \ \ \ +kh\frac{\lambda+\mu}{\lambda+2\mu}(\alpha_m e^{-kh}-\beta_m e^{kh})\bigg]=0\ .\nonumber\\
\label{eq:BC4}
\end{eqnarray}
Boundary condition (v) reads
\begin{eqnarray}
(a_2+2b_2)a=0.\label{eq:BC5}
\end{eqnarray}
Boundary condition (vi) will be almost satisfied by minimizing $I\equiv\int_0^hT_{rr}^2(z)dz$.
\begin{eqnarray}
T_{rr}&\equiv&\Theta(z)+c_0+c_1z\ ,\\
\Theta&=&\left[(\lambda+2\mu)kA_m+\lambda B_m'\right]J_0(ka)\ ,\\
c_0&=&2(\lambda+\mu)a_1+\lambda b_3\ ,\label{eq:a1}\\
c_1&=&2(\lambda+\mu)a_2+2\lambda b_1\ .\label{eq:a2}
\end{eqnarray}
From Eqs.~(\ref{eq:BC1})-(\ref{eq:BC4}) we obtain

\onecolumngrid
\begin{eqnarray}
\alpha_m&=&\frac{F_0p_m(\lambda+2\mu)}{k\mu(\lambda+\mu)}\frac{\epsilon_1(1-Q+2khQ)+\epsilon_2\sqrt{Q}(1+2kh-Q)}{(1-Q)^2-4k^2h^2Q}\ ,\nonumber\\
\beta_m&=&\frac{F_0p_m(\lambda+2\mu)Q}{k\mu(\lambda+\mu)}\frac{\epsilon_1(1-Q+2kh)+\epsilon_2/\sqrt{Q}(1+2khQ-Q)}{(1-Q)^2-4k^2h^2Q}\ ,\nonumber\\
\gamma_m&=&-\frac{F_0p_m}{2k\mu(\lambda+\mu)}\frac{\epsilon_1([2k^2h^2(\lambda+\mu)+2\mu kh]Q+\mu(1-Q))+\epsilon_2\sqrt{Q}(\mu(1-Q)+kh[(\lambda+\mu)(1-Q)+2\mu])}{(1-Q)^2-4k^2h^2Q}\ ,\nonumber\\
\delta_m&=&-\frac{F_0p_mQ}{2k\mu(\lambda+\mu)}\frac{\epsilon_1([2k^2h^2(\lambda+\mu)-2\mu kh]-\mu(1-Q))+\epsilon_2/\sqrt{Q}(-\mu(1-Q)+kh[(\lambda+\mu)(1-Q)-2\mu Q])}{(1-Q)^2-4k^2h^2Q}\nonumber\\
\end{eqnarray}

\twocolumngrid
\ \\
where $Q=\exp[-2kh]$. Also $b_1$, $b_2$, $b_3$ can be expressed by $a_1$ and $a_2$ as:
\begin{eqnarray}
b_1&=&-\frac{\lambda}{\lambda+2\mu}a_2+\frac{1}{2(\lambda+2\mu)}(\epsilon_1+\epsilon_2)\frac{F_0p_0}{h}\ ,\nonumber\\
b_2&=&-\frac{a_2}{2}\ ,\\
b_3&=&-\frac{2\lambda}{\lambda+2\mu}a_1-\frac{1}{\lambda+2\mu}\epsilon_1F_0p_0\ .\nonumber
\end{eqnarray}
Now let us take the derivative of $I$:
\begin{eqnarray}
\frac{\partial I}{\partial c_0}=0&\rightarrow&c_0=-\frac{1}{h}\int_0^h\Theta(z)dz-\frac{h}{2}c_1\ ,\label{eq:c0}\\
\frac{\partial I}{\partial c_1}=0&\rightarrow&c_1=-\frac{3}{h^3}\int_0^h\Theta(z)zdz-\frac{3}{2h}c_0\ .\nonumber\\ \label{eq:c1}
\end{eqnarray}
After some algebra with the boundary conditions (i)-(iv), we obtain
\begin{eqnarray}
\int_0^h\Theta(z)dz&=&0\ ,\\
\int_0^h\Theta(z)zdz&=&
\sum_m\frac{\epsilon_1+\epsilon_2}{k^2}F_0p_mJ_0(ka)\ .\nonumber\\
\end{eqnarray}
Combining these with Eqs.~(\ref{eq:c0}) and (\ref{eq:c1}), we have
\begin{eqnarray}
c_0&=&\sum_m\frac{6(\epsilon_1+\epsilon_2)}{k^2h^2}F_0p_mJ_0(ka)\ ,\\
c_1&=&-\sum_m\frac{12(\epsilon_1+\epsilon_2)}{k^2h^3}F_0p_mJ_0(ka)\ ,
\end{eqnarray}
and then $a_1$ and $a_2$ are given from Eqs.~(\ref{eq:a1}) and (\ref{eq:a2}). Finally, we obtain

\onecolumngrid

\begin{eqnarray}
\delta u_r&=&\frac{\lambda+2\mu}{2\mu(3\lambda+2\mu)}(c_0r+c_1rz)+\frac{\lambda F_0p_0r}{2\mu(3\lambda+2\mu)}\left\{\epsilon_1-(\epsilon_1+\epsilon_2)\frac{z}{h}\right\}\ ,\\
\delta u_z&=&-\frac{\lambda}{\mu(3\lambda+2\mu)}\left(c_0z+\frac{c_1z^2}{2}\right)-\frac{\lambda+2\mu}{4\mu(3\lambda+2\mu)}c_1r^2\nonumber\\
&&\hspace{1cm}-\frac{(\lambda+\mu)F_0p_0}{\mu(3\lambda+2\mu)}\left\{\epsilon_1z-(\epsilon_1+\epsilon_2)\frac{z^2}{2h}\right\}+\frac{\lambda F_0p_0(\epsilon_1+\epsilon_2)r^2}{4\mu(3\lambda+2\mu)h}\ .\label{eq:samebyhere}
\end{eqnarray}

\twocolumngrid
As is introduced by Harry {\it et al.}~\cite{Harry}, the boundary conditions between the substrate and the front-surface coatings are:
\begin{align}
E'_{rr}(r)            &=E_{rr}(r, 0)                 \ ,\nonumber\\
E'_{\psi\psi}(r)&=E_{\psi\psi}(r, 0)  \ ,\nonumber\\
E'_{rz}(r)            &=E_{rz}(r, 0) =0            \ ,\nonumber \\
T'_{rz}(r)            &=T_{rz}(r, 0) =0            \ ,\nonumber \\
T'_{zz}(r)            &=T_{zz}(r, 0)              \ , \label{eq:boundary1}
\end{align}
where the elements with a prime ($'$) are for the front-surface coatings. Since our probe forces are oriented along the normal of the cylinder's flat faces (i.e. z-axis) they induce no shear to the etalon. Consequently, at the boundary and in the coatings $E_{rz}$, $E'_{rz}$, $T_{rz}$, and $T'_{rz}$ are zero. Since the coatings are thin, we can assume that the strain and stress tensor elements are constant in terms of $z$. The strain tensor elements of the front-surface coatings are then given as

\onecolumngrid

\begin{eqnarray}
&&E'_{rr}=\sum_m{A_m(0)\frac{k}{2}\Bigl[J_0(kr)-J_2(kr)\Bigr]}+\frac{\lambda+2\mu}{2\mu(3\lambda+2\mu)}c_0+\frac{\lambda \epsilon_1 F_0p_0}{2\mu(3\lambda+2\mu)}\ ,\nonumber\\
&&E'_{\psi\psi}=\sum_m{A_m(0)\frac{k}{2}\Bigl[J_0(kr)+J_2(kr)\Bigr]}+\frac{\lambda+2\mu}{2\mu(3\lambda+2\mu)}c_0+\frac{\lambda \epsilon_1 F_0p_0}{2\mu(3\lambda+2\mu)}\ ,\nonumber\\
&&E'_{zz}=\frac{-\lambda'}{\lambda'+2\mu'}\left[\sum_m{kA_m(0)J_0(kr)}+\frac{\lambda+2\mu}{\mu(3\lambda+2\mu)}c_0+\frac{\lambda\epsilon_1 F_0p_0}{\mu(3\lambda+2\mu)}\right]\nonumber\\
&&\hspace{1.5cm}-\frac{1}{\lambda'+2\mu'}\left(\sum_m{\epsilon_1F_0p_mJ_0(kr)}+\epsilon_1 F_0p_0\right)\ ,\label{eq:strain1}
\end{eqnarray}

\twocolumngrid
\ \\
and the stress tensor elements of the coatings are given as
\begin{eqnarray}
T'_{rr}&=&(\lambda'+2\mu')E'_{rr}+\lambda'(E'_{\psi\psi}+E'_{zz})\ ,\nonumber\\
T'_{\psi\psi}&=&(\lambda'+2\mu')E'_{\psi\psi}+\lambda'(E'_{zz}+E'_{rr})\ ,\nonumber\\
T'_{zz}&=&(\lambda'+2\mu')E'_{zz}+\lambda'(E'_{rr}+E'_{\psi\psi})\ .\label{eq:stress1}
\end{eqnarray}
Putting these into
\begin{eqnarray}
U'=\pi d_1\int_0^a\!\sum_j{E'_{jj}T'_{jj}}rdr\ \ (j=r,\psi,z)\ ,\label{eq:U2}
\end{eqnarray}
and then into Eq.~(\ref{eq:FdT}), we obtain the power spectrum of thermal noise originating from the loss in the front-surface coatings.

The boundary conditions between the substrate and the back-surface coatings are:
\begin{align}
E''_{rr}(r)                &=E_{rr}(r, h)                 \ ,\nonumber \\
E''_{\psi\psi}(r)    &=E_{\psi\psi}(r, h)     \ ,\nonumber\\
E''_{rz}(r)                &=E_{rz}(r, h)=0             \ , \nonumber\\
T''_{rz}(r)                &=T_{rz}(r, h)=0             \ ,\nonumber\\
T''_{zz}(r)                &=T_{zz}(r, h)-\epsilon_2F_0p(r)=0 \ ,\label{eq:boundary2}
\end{align}
where the elements with a double prime ($''$) are for the back-surface coatings. Note that $E''_{rz}$, $T''_{rz}$, and $T''_{zz}$ are zero. Since the light on the back surface is reflected back at the boundary of the substrate and the coatings, the imaginary force $-\epsilon_2F_0p(r)$ is applied at the boundary. The stress in the $z$-direction in the back-surface coatings is then zero. This means that the expansion of the back-surface coatings does not appear in coating thermal noise. The strain tensor elements of the back-surface coatings are given as
\onecolumngrid

\begin{eqnarray}
&&E''_{rr}=\sum_m{A_m(h)\frac{k}{2}\Bigl[J_0(kr)-J_2(kr)\Bigr]}+\frac{\lambda+2\mu}{2\mu(3\lambda+2\mu)}(c_0+c_1h)-\frac{\lambda \epsilon_2 F_0p_0}{2\mu(3\lambda+2\mu)}\ ,\nonumber\\
&&E''_{\psi\psi}=\sum_m{A_m(h)\frac{k}{2}\Bigl[J_0(kr)+J_2(kr)\Bigr]}+\frac{\lambda+2\mu}{2\mu(3\lambda+2\mu)}(c_0+c_1h)-\frac{\lambda \epsilon_2 F_0p_0}{2\mu(3\lambda+2\mu)}\ ,\nonumber\\
&&E''_{zz}=\frac{-\lambda''}{\lambda''+2\mu''}\sum_m{kA_m(h)J_0(kr)}-\frac{\lambda''\left[(\lambda+2\mu)(c_0+c_1h)-\lambda\epsilon_2 F_0p_0\right]}{(\lambda''+2\mu'')\mu(3\lambda+2\mu)}\ ,\label{eq:strain2}
\end{eqnarray}
\twocolumngrid
\ \\
and the stress tensor elements of the coatings are given in the same way as in Eq.~(\ref{eq:stress1}). Putting these into
\begin{eqnarray}
U''=\pi d_2\int_0^a\!\sum_j{E''_{jj}T''_{jj}}rdr\ \ (j=r,\psi,z)\ ,\label{eq:U''2}
\end{eqnarray}
and then into Eq.~(\ref{eq:FdT}), we obtain the power spectrum of thermal noise originated from the loss in the back-surface coatings. 

\begin{figure}[htbp]
  \centering
    \includegraphics[width=6cm]{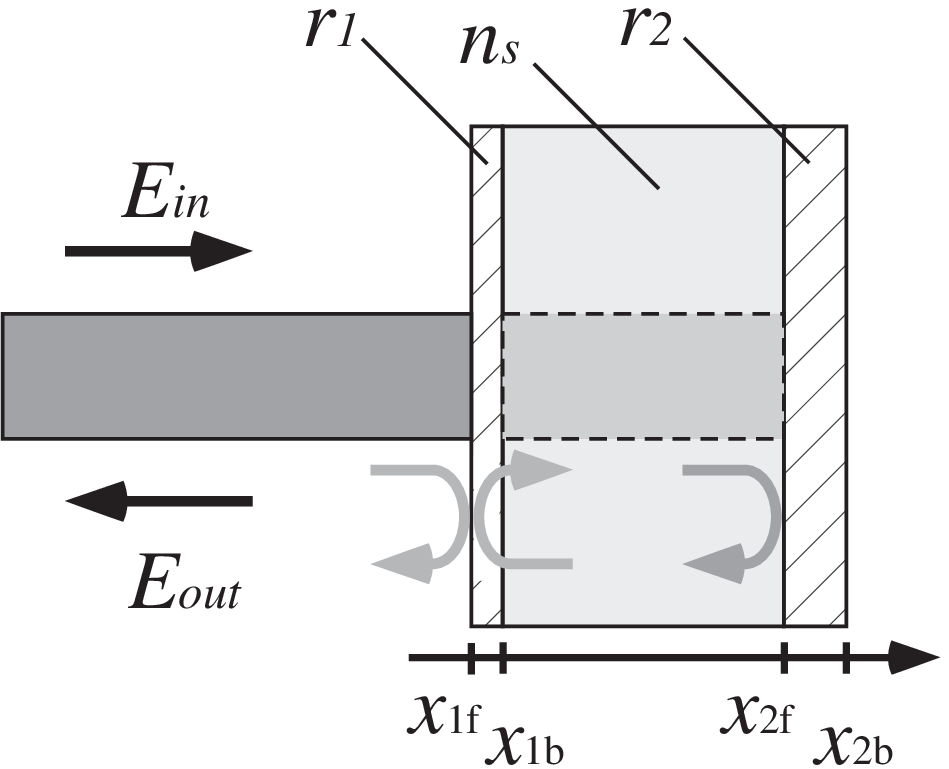} \\
\vspace{0.5cm}
    \includegraphics[width=6cm]{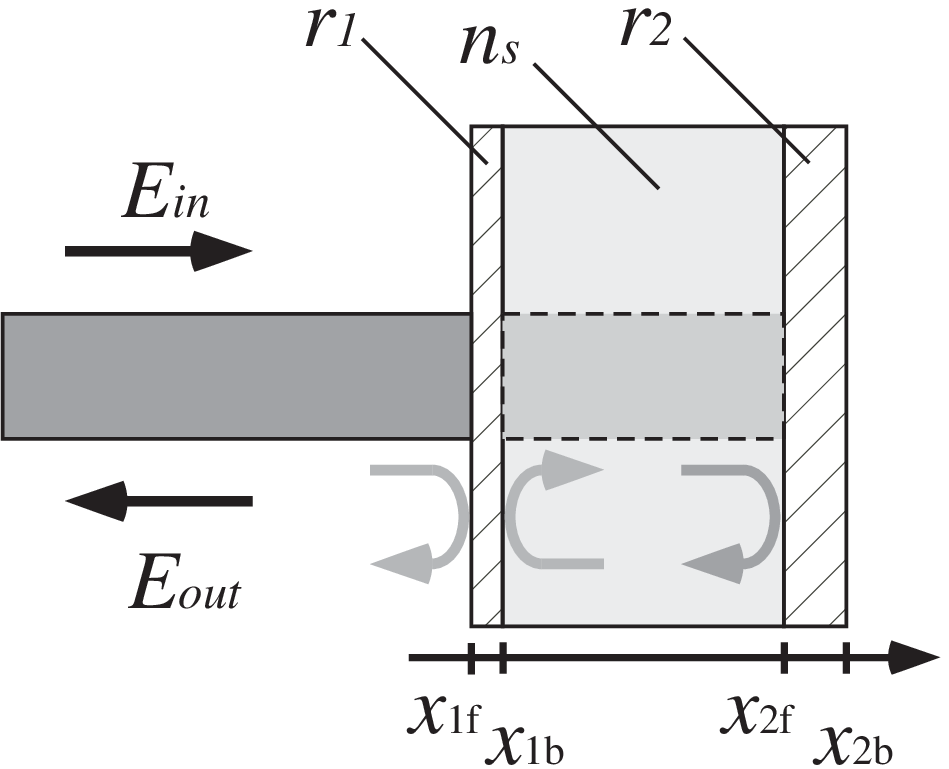}
  \caption{{\it Top} : Thin-layer model. The light at the front surface is always reflected on the front side of the coatings. {\it Bottom} : Thick-layer model. The light circulating inside the etalon is reflected on the rear side of the front surface coatings while the incident light is reflected on the front side of the coatings. The mathematical formulation with the thin-layer model is more simple while the thick-layer model is more realistic.}
\label{fig:thick}
\end{figure}

\section{Thick-layer model}
\label{sec:thickLayer}

In the thick-layer model, we take into account the fact that the light circulating in the etalon probes the rear side of the front coating. Figure~\ref{fig:thick} explains the difference between the two models. Imaginary forces are applied at three different planes in the thick-layer model since the reflection planes are actually different for the incident light and the light circulating in the etalon substrate. In fact, since there are anyway two boundary conditions on $T_{zz}$, the complexity to solve the elastic equation does not increase much. We apply the imaginary force as usual on the rear side of the front surface coatings, namely the interface between the coating layers and the substrate. It is the stress tensor inside the front surface coatings that makes a slight difference from the thin-layer model.

In the thin-layer model, we applied the imaginary force with the weight function $\epsilon_1$ on the front side of the front surface coatings, which imposes thermal noise due to the expansion of the coatings in the $z$-direction by $\epsilon_1$. If we apply the imaginary force on the rear side of the front surface coatings, the expansion is not probed by the light, which is the case for the back surface coatings. What happens in reality in the front surface coatings is intermediate of the two extreme situations. A fraction of the light is reflected on the front side and probes the expansion, and the rest transmits through the coatings to be reflected by the back surface coatings of the etalon and nevertheless probes the expansion but in a different way due to the difference of the refraction index.

\subsection{Probe force amplitudes $\epsilon_j$}

The optical path length between the rear side of the front surface and the front side of the back surface is kept an odd multiple of the wavelength in the zeroth order. The optical thickness of the front surface coatings is also kept an odd multiple of the wavelength in the zeroth order. Phase of the light reflected by the etalon changes according to the motion of each boundary $\delta x_\mathrm{1f}$, $\delta x_\mathrm{1b}$, and $\delta x_\mathrm{2f}$. The reflected light field $E_{out}$ is given with the input field $E_{in}$ as
\begin{eqnarray}
\frac{E_{out}}{E_{in}}&=&-e^{2ik_0\delta x_\mathrm{1f}} \left( r_1 - e^{2i\theta_c} \frac{(1-r_1^2) r_2 e^{2i\theta_s}}{1-r_1r_2e^{{2i\theta_s}}} \right)\ ,\nonumber\\
\theta_s&=&k_0n_s(\delta x_\mathrm{2f}-\delta x_\mathrm{1b})\ , \nonumber \\
\theta_c&=&k_0n_c(\delta x_\mathrm{1b}-\delta x_\mathrm{1f})\ ,\label{eq:BandA2}
\end{eqnarray} 
where $n_c$ is the refractive index of the coating layers; the index for silica to calculate thermal noise of silica coatings and the index for tantala to calculate thermal noise of tantala coatings. Expanding Eq.~(\ref{eq:BandA2}) into a series over $\delta x_\mathrm{1f}$, $\delta x_\mathrm{1b}$, and $\delta x_\mathrm{2f}$, and keeping linear terms, we obtain
\begin{eqnarray}
\frac{E_{out}}{E_{in}}\simeq-\frac{r_1+r_2}{1+r_1r_2}-2ik_0\left( \epsilon_\mathrm{1f}\delta x_\mathrm{1f} + \epsilon_\mathrm{1b}\delta x_\mathrm{1b} +\epsilon_\mathrm{2f}\delta x_\mathrm{2f} \right)\ ,\nonumber\\
\end{eqnarray}
with
\begin{eqnarray}
\epsilon_\mathrm{1f}&=&\frac{r_1+r_2}{1+r_1r_2}-\frac{n_c(1-r_1^2)r_2}{1+r_1r_2} \nonumber\\ 
\epsilon_\mathrm{1b}&=&\frac{n_c(1-r_1^2)r_2}{1+r_1r_2}-\frac{n_s(1-r_1^2)r_2}{(1+r_1r_2)^2} \nonumber\\ 
\epsilon_\mathrm{2f}&=&\frac{n_s(1-r_1^2)r_2}{(1+r_1r_2)^2}  \ .\label{eq:epsilon2}
\end{eqnarray}
These $\epsilon_j$ are combined to be used as weight coefficients of the imaginary forces applying to the surfaces of the etalon. Comparing Eqs.~(\ref{eq:epsilon}) and (\ref{eq:epsilon2}), we can see $\epsilon_\mathrm{1f}+\epsilon_\mathrm{1b}=\epsilon_1$ and $\epsilon_\mathrm{2f}=\epsilon_2$. Note that $\epsilon_\mathrm{1f}$ would simply become $r_1$ if $n_c=1$, which represents the fact that the light transmitting through the front surface coatings also probes the expansion of the coatings due to the difference of the refraction index of the coatings from that of the vacuum.

\subsection{Elastic motion of the substrate}

Most part of the calculation process for the thick-layer model is common to the process for the thin-layer model that we have shown in Sec.~\ref{subsec:thinlayerelastic}. First we solve the elastic equation of the etalon substrate and then extend the solution to the coatings. There are three $\epsilon_j$ coefficients in the thick-layer model, but we need only two weighting coefficients to be multiplied to the imaginary forces on the front and back surfaces of the etalon substrate, and as a result the weighting coefficients are the same as those in the thin-layer model so that the solution of the elastic equation is exactly the same. It is only the expansion of the front surface coatings that makes a difference between the two models.

In using the Fluctuation-dissipation theorem, the thick-layer model says that the imaginary force $\epsilon_\mathrm{1f}F_0$ is applied on the front surface coatings and the imaginary force $\epsilon_\mathrm{1b}F_0$ is applied directly to the substrate. However, due to the thinness of the coatings (even in the thick-layer model), the force applied on the coatings is directly transferred to the substrate. The total force applied to the substrate is therefore $\epsilon_\mathrm{1f}F_0+\epsilon_\mathrm{1b}F_0=\epsilon_1F_0$, which is exactly the same as the imaginary force on the front surface coatings in the thin-layer model. The force applied to the back surface is apparently $\epsilon_\mathrm{2f}F_0=\epsilon_2F_0$.

The calculation process is common up to Eq.~(\ref{eq:samebyhere}), but one of the boundary conditions (\ref{eq:boundary1}) becomes different:
\begin{eqnarray}
\left\{
\begin{array}{@{\,}l@{\,}}
\mathrm{[thin\ layer\ model]}\\
\hspace{0.5cm}T'_{zz}(r)=T_{zz}(r,\ 0)=-\epsilon_1F_0p(r)\\
\mathrm{[thick\ layer\ model]}\\
\hspace{0.5cm}T'_{zz}(r)=T_{zz}(r,\ 0)+\epsilon_\mathrm{1b}F_0p(r)=-\epsilon_\mathrm{1f}F_0p(r)
\end{array}\right.
\end{eqnarray}
which tells us that the expansion of the coatings is in most cases slightly overestimated in the thin-layer model. Accordingly, the stress tensor element $E_{zz}$, the last line of Eq.~(\ref{eq:strain1}), becomes different:
\begin{eqnarray}
&&\mathrm{[thick\ layer\ model]}\nonumber\\
&&E'_{zz}=\frac{-\lambda'}{\lambda'+2\mu'}\left[\sum_m{kA_m(0)J_0(kr)}\right.\nonumber\\
&&\hspace{1cm}\left.+\frac{\lambda+2\mu}{\mu(3\lambda+2\mu)}c_0+\frac{\lambda\epsilon_1 F_0p_0}{\mu(3\lambda+2\mu)}\right]\\
&&\hspace{1.2cm}-\frac{1}{\lambda'+2\mu'}\left(\sum_m{\epsilon_\mathrm{1f}F_0p_mJ_0(kr)}+\epsilon_\mathrm{1f} F_0p_0\right)\ .\nonumber
\end{eqnarray}

\section{Results}
\label{sec:results}

According to the non-zero correlation between the motions of the front and back surfaces, the power spectrum of total coating thermal noise contains not only the power spectrum of the motion of each surface, $S_1(\Omega)$ and $S_2(\Omega)$, but also the cross spectrum $S_{12}(\Omega)$:
\begin{eqnarray}
S_x(\Omega)&=&\epsilon_1^2 S_1+\epsilon_2^2 S_2+2\epsilon_1\epsilon_2 S_{12}\label{eq:twopaths}\ .
\end{eqnarray}
However, if we use $\tilde{S}_1$ and $\tilde{S}_2$, which are respectively the power spectrum of noise caused by the thermal energy in the front surface and back surface coatings, the total etalon noise spectrum $S_x$ can be expressed without the cross spectrum as
\begin{multline}
S_x(\Omega)=
\left(\epsilon_1^2+\eta^2 \epsilon_2^2+2\chi_1\,\epsilon_1\,\eta\epsilon_2\right)\tilde{S}_{1} \\
+\left(\eta^2\epsilon_1^2+\zeta^2\epsilon_2^2+2\chi_2\,\eta\epsilon_1\,\zeta\epsilon_2\right)\tilde{S}_{2} \ .
\label{eq:tot_noise}
\end{multline} 
While the thermal motions of the surfaces can be correlated, the origins of the fluctuations must be independent. Here the coefficient $\eta$ represents the mechanical transfer from one surface to the other. The coefficient $\zeta$ results from the difference of the application points of the probe force, and the value is slightly different between the thin-layer model and the thick-layer models. With our parameters, however, $\zeta$ is about 80~\% in either model. The mechanical transfer $\eta$ can be obtained via noise calculations with lossless back-surface coatings. In this case it equals the noise ratio, when the same force is applied separately onto the back and the front of the etalon. If we repeat this calculation with lossless front-surface coatings, we obtain the slightly different expression $\eta/\zeta$, showing the breaking of symmetry in our model. The coefficients $\chi_1$ and $\chi_2$ represent the correlation of the motion generated from the same dissipation source but appearing on the different surfaces. See App.~\ref{app:mechtrans} for more details.

\begin{figure}[thbp!]
      \begin{center}
		\includegraphics[width=8cm]{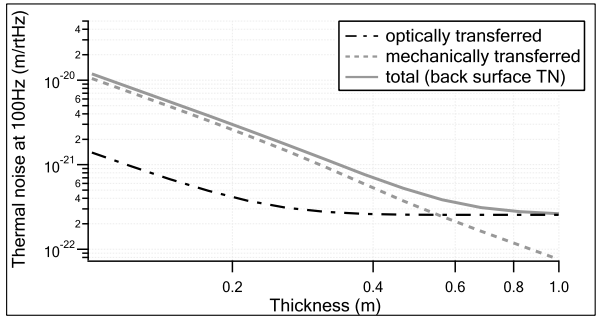}
	\caption{Thermal noise originating from the back-surface coatings appears via two different paths. (i) The motion is measured by the probe on the back surface (optically transferred). (ii) The motion transfers to the front surface through the substrate and is measured by the probe on the front surface (mechanically transferred).}
	\label{fig:epsilon}
      \end{center}
\end{figure}

The comparison of the first two terms for the back-surface coatings, i.e. the $\eta\epsilon_1$-term and the $\zeta\epsilon_2$-term, is shown in Fig.~\ref{fig:epsilon}. Here we assume the proposed end mirror of the Einstein Telescope (see Table~\ref{table:ET}).  The front surface has 3 doublets and the back surface has 17 doublets of coating layers, which makes the reflectivity of the etalon as high as that of a single mirror with 19 doublets; "$N$ doublets" means that there are $N$ quarter-wavelength layers of tantala, $N-1$ quarter-wavelength layers of silica, and one half-wavelength silica cap layer. For the assumed mirror thickness of 30~cm, the mechanically transferred motion ($\eta\epsilon_1$-term) is larger than the optically transferred motion ($\zeta\epsilon_2$-term). The measurement frequency is fixed at 100~Hz.

\begin{figure}[thbp!]
      \begin{center}
		\includegraphics[width=8cm]{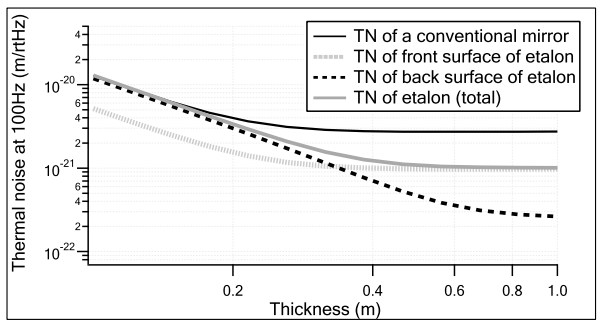}
	\caption{Thermal noise of etalon depends on the thickness of the mirror. The total noise level of the etalon can be a factor of $\sim1.7$ times better than that of the single mirror with the same reflectivity when the mirror thickness is 30~cm.}
	\label{fig:TN}
      \end{center}
\end{figure}

Figure~\ref{fig:TN} shows the total thermal-noise level of the anti-resonant etalon. Here we use the thick-layer model to calculate the thermal-noise level, but the difference between the models is less than 1~\%. The noise level of the etalon is a factor of $\sim1.7$ times better than that of the single mirror with the same reflectivity. We could reduce the noise level further by using a thicker etalon. In this case, thermorefractive noise~\cite{TRc} will increase, but it turns out to be a few orders of magnitude smaller than coating thermal noise with our parameters~\cite{AlexeyKE}. Increasing the thickness to 40~cm, we can further reduce the noise level by $\sim37~\%$.

\begin{table*}
\begin{center}
\begin{tabular}{|lr|}
\hline
mirror radius&31~cm\\
mirror thickness&30~cm\\
beam radius&12~cm\\
\hline
$\phi$ of silica layers&$4\times10^{-5}$\\
$\phi$ of tantala layers&$2\times10^{-4}$\\
\hline
number of silica layers$^{(*)}$&2 on front, 16 on back\\
number of tantala layers&3 on front, 17 on back\\
amplitude reflectivity of the front coating & $r_1=0.834$\\
\hline
amplitude reflectivity of the back coating & $r_2=1-1.36\times 10^{-5}$\\
force coefficient of the front &$\epsilon_1=0.869$ \\
force coefficient of the back & $\epsilon_2=0.131$ \\
\hline
Young's modulus (silica)&$7.2\times10^{10}$~Pa\\
Young's modulus (tantala)&$1.4\times10^{11}$~Pa\\
Poisson ratio (silica)&0.17\\
\hline
Poisson ratio (tantala)&0.23\\
refractive index (silica)&1.45\\
refractive index (tantala)&2.035\\
\hline
temperature&290~K\\
Boltzmann constant&$1.38\times10^{-23}$~J/K\\
wavelength of light &1064~nm \\
\hline
\end{tabular}
\caption{Parameters used in this paper. $^{(*)}$: A half-wavelength silica layer was placed on top of these quarter-wavelength doublets for protection.}
\label{table:ET}
\end{center}
\end{table*}

\section{Discussions}
\label{sec:discussion}
\subsection{Cross check with FEA}

In order to verify the results derived in the last section, we have performed the numerical calculation using a finite element analysis (FEA). With the help of the COMSOL program package~\cite{comsol} we applied two virtual Gaussian pressures on the front and back surface and calculated the elastic response, i.e. the strain energy distribution. In accordance with the thin layer model of Sec.~\ref{sec:thinLayer}, firstly the elastic response of the substrate without any coatings was determined. Due to the small fraction of coating compared to the substrate, this approximation will be applicable. We carefully computed the error due to this approximation to be well below one percent. Secondly the strain energy in the coatings results from the consideration of the boundary conditions between substrate and coating as is shown in Eqs.~(\ref{eq:boundary1}) and (\ref{eq:boundary2}). The numerical code was successfully tested on the analytical model of Brownian bulk noise given by Liu and Thorne \cite{Liu}. Thus, it can be regarded as an independent tool to confirm the semi-analytical derivation.

We applied both results on a fused silica test mass characterized in table~\ref{table:ET}. A comparison of the elastic energy density in the different coatings is given in table~\ref{tab:comsol}. Both the analytical and FEA results match within 5~\%. Due to the Fluctuation-dissipation-theorem a coincidence of dissipated energy will lead to the same level of Brownian noise for both approaches. These results clearly confirm the analytical calculations.
\begin{table*}
\begin{center}
\begin{tabular}{|c|c|c|c|}
\hline
	position & material & U [J/m] FEA & U [J/m] analytical \\
	\hline
	front	&	silica  & 19.0 $\times$ 10$^{-11}$& 19.1 $\times$ 10$^{-11}$ \\
	front	& tantala & 21.7 $\times$ 10$^{-11}$& 21.9 $\times$ 10$^{-11}$ \\
	back	&	silica  & 3.15 $\times$ 10$^{-11}$& 3.25 $\times$ 10$^{-11}$ \\
	back	& tantala & 6.69 $\times$ 10$^{-11}$& 6.64 $\times$ 10$^{-11}$ \\
	\hline
\end{tabular}
\caption{Energy density stored in the coating layers of a Khalili cavity in comparison to the analytical results at 100~Hz.}
\label{tab:comsol}
\end{center}
\end{table*}

\begin{figure}[htbp!]
      \begin{center}
		\includegraphics[width=7cm]{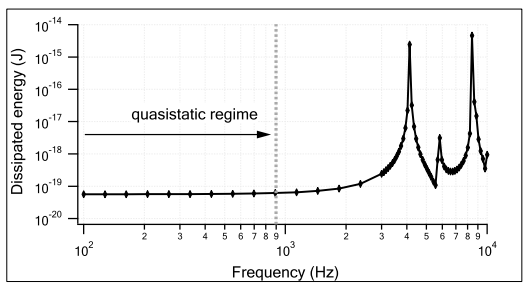}
        \caption{Frequency dependence of coating dissipation for Brownian noise. The mirror parameters are shown in Table~\ref{table:ET}.}
        \label{fig:comsol}
      \end{center}
\end{figure}

Additionally the FEA calculation provides a good measure to estimate the accuracy of the quasistatic analytical calculation. Increasing the vibrational frequency will lead to a failure of the quasistatic approach, typically starting around the lowest resonant frequencies of the test mass. With the help of the sample's layer stack geometry one can calculate the total dissipated energy corresponding to the expected Brownian noise. Numerical results of the frequency behavior of the total dissipated energy are given in Fig.~\ref{fig:comsol}. They show a deviation from the quasistatic model of less than 10~\% for frequencies lower than 900~Hz.

\subsection{Analytical cross check on the thin mirror limit}

Another way to verify the result in a more intuitive way is to compare the thermal-noise level when the mirror is so thin that the mechanical transfer from one surface to the other is almost direct (i.e. $\eta\simeq1$).

The strain and stress tensors are described by the sum of the Bessel's functions and zeroth- and first-order polynomials of $r$ and $z$. The coefficients of the functions and the polynomials are given by the combination of $\alpha_m$, $\beta_m$, $\gamma_m$, $\delta_m$, $c_0$, and $c_1$. Using an approximation $h\ll a$ (or $kh\ll1$), these coefficients become
\begin{eqnarray}
\alpha_m&\sim&\frac{Fp_m(\lambda+2\mu)}{k\mu(1+\mu)}\frac{(\epsilon_1+\epsilon_2)}{2kh}\ ,\\
\beta_m&\sim&\frac{Fp_m(\lambda+2\mu)}{k\mu(1+\mu)}\frac{(\epsilon_1+\epsilon_2)}{2kh}\ ,\\
\gamma_m&\sim&-\frac{Fp_m}{2k(\lambda+\mu)}\frac{(\epsilon_1+\epsilon_2)}{2kh}\ ,\\
\delta_m&\sim&-\frac{Fp_m}{2k(\lambda+\mu)}\frac{(\epsilon_1+\epsilon_2)}{2kh}\ ,\\
c_0&\sim&\frac{6a^2}{h^2}(\epsilon_1+\epsilon_2)\sum_m{\frac{J_0(ka)p_m}{ka}}\ ,\\
c_1&\sim&-\frac{12a^2}{h^3}(\epsilon_1+\epsilon_2)\sum_m{\frac{J_0(ka)p_m}{ka}}\ .
\end{eqnarray}
One can see that $\epsilon_1$ and $\epsilon_2$ appear in the coefficients in the form of $\epsilon_1+\epsilon_2$, which is actually the reflectivity of the system:
\begin{eqnarray}
\epsilon_1+\epsilon_2=\frac{r_1+r_2}{1+r_1r_2}\simeq1\ .
\end{eqnarray}
The thermal-noise level of a conventional mirror is given by the same equations with $r_1\simeq1$ and $r_2=0$. The results of this case match with a former calculation of thermal coating noise of a conventional mirror~\cite{SomiyaTN}.

\subsection{Possible noise cancellation in a resonant Khalili etalon}

\label{sec:cancellation}
In the case of the etalon in this paper, noise contributions from different dissipation sources (front- or back-surface coatings) through different paths (optical and mechanical transfer) add up to make total noise. If the light is not on anti-resonance in the etalon like in our case but on resonance in the etalon, the two noise contributions will not add up but partially compensate so that the total noise level could be lower than the anti-resonant etalon. The $\epsilon_j$ terms become
\begin{eqnarray}
\epsilon^\mathrm{reso}_\mathrm{1f}&=&\frac{r_1-r_2}{1-r_1r_2}+\frac{n_c(1-r_1^2)r_2}{1-r_1r_2}\ ,\nonumber\\ 
\epsilon^\mathrm{reso}_\mathrm{1b}&=&-\frac{n_c(1-r_1^2)r_2}{1-r_1r_2}+\frac{n_s(1-r_1^2)r_2}{(1-r_1r_2)^2}\ ,\nonumber\\ 
\epsilon^\mathrm{reso}_\mathrm{2f}&=&-\frac{n_s(1-r_1^2)r_2}{(1-r_1r_2)^2}\ ,
\end{eqnarray}
when the light is on resonance in the etalon. The reflectivity of the etalon is equal to $\epsilon^\mathrm{reso}_\mathrm{1f}+\epsilon^\mathrm{reso}_\mathrm{1b}+\epsilon^\mathrm{reso}_2=(r_1-r_2)/(1-r_1r_2)$, which is lower than $\epsilon_1+\epsilon_2$. With the parameters we used in the paper, the power transmittance of the resonant etalon is 91~ppm while that of the anti-resonant etalon is 2.5~ppm. Note that the thick layer model is suitable in the case of the resonant etalon.

\begin{figure}[htbp!]
      \begin{center}
		\includegraphics[width=7cm]{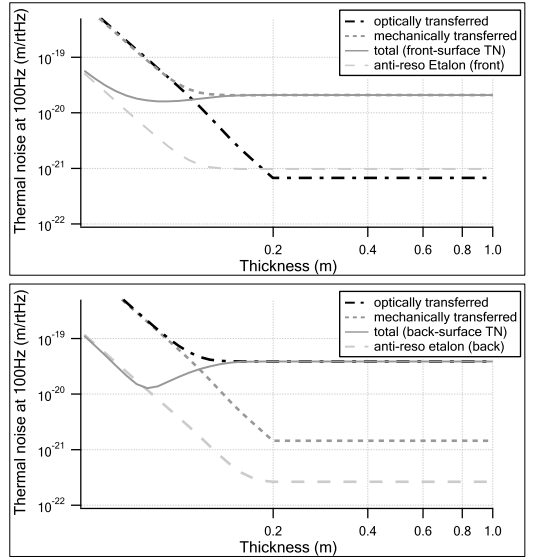}
        \caption{Thermal noise of an etalon with the light being resonant in the substrate. Noise contribution from the dissipation caused in the front-surface coatings (top) and in the back-surface coatings (bottom). Part of the motions probed on the front and the back surfaces partially compensate but the total noise level is higher than the anti-resonant etalon.}
        \label{fig:cancellation}
      \end{center}
\end{figure}

Figure~\ref{fig:cancellation} shows the noise level of the resonant etalon. In the top panel, the origin of the fluctuation is the front-surface coatings, and in the bottom panel, the origin is the back-surface coatings. In each panel, the total noise level of the resonant etalon is given by the coherent sum of the mechanically transferred motion and the optically transferred motion. Since the fluctuations through different paths partially compensate, the total noise level is lower than the noise level of the fluctuation through a single path. Due to the fact that the light circulates and probes the fluctuations more in the resonant etalon, however, the noise level is higher than that in the anti-resonant etalon in most cases. There could be a better parameter set to realize more significant improvement by the compensation, but we shall leave this for a future work.

\subsection{Multi layer model}\label{sec:multi}

In this paper, we approximated the multi-layer coatings as a single thick layer of silica or tantala. This mono-layer approximation is used in Ref.~{\cite{SomiyaTN}\cite{Harry}}. Reference~\cite{SomiyaMonoLayer} explains that the stress tensor and the strain tensor of each layer are given by those of interface between the substrate and the neighboring layer if we assume the stress is equal within each thin layer, thus the square-sum of the elastic energy in each layer is equal to that of a single thick layer. This approach is based on the assumption that the light reflects on the first surface of the coatings and the light propagates without reflections in the coatings. While this assumption is a good approximation, the light is actually reflected on each boundary between the layers. Each displacement $\delta x_k$ of the $k$-th boundary of the layers contributes to the phase shift of the reflected light. The total displacement $\delta x$ read out by the reflected beam is given as
\begin{eqnarray}
\label{dxi}
 \delta x =\sum_k\epsilon_k \delta x_k\ ,
\end{eqnarray}
where $\epsilon_k$ is the coefficient that represents a contribution of each boundary. With this approach, recently demonstrated in Ref.~\cite{Alexey}, the power spectrum of coating thermal noise for a conventional mirror can be corrected by a few percent. In the case of an etalon, the correction of the power spectrum is more than the conventional mirror:
\begin{eqnarray}
\frac{S_\mathrm{mono}(\Omega)-S_\mathrm{multi}(\Omega)}{S_\mathrm{multi}(\Omega)}\simeq 0.10
\end{eqnarray}
where $S_\mathrm{mono}(\Omega)$ is calculated with the mono-layer approximation (used in this paper) and $S_\mathrm{multi}(\Omega)$ is calculated with this approach with multi-layers.

\section{Summary}

In this paper we have shown a semi-analytical method to calculate coating thermal noise of an etalon, which can be used to lower the total thermal-noise level with the same reflectivity as a conventional mirror if we put a few coatings on the front surface and more on the back surface ({\it Khalili etalon}). We used the Fluctuation-dissipation theorem with probes on both sides of the etalon so that the thermal motion due to the same dissipation source can be summed up coherently. The result revealed a strong dependence of the noise contribution on the thickness of the etalon. In fact, the noise level of the etalon can be as low as that of a separate 2-mirror system, if the etalon is sufficiently thick. The result was verified with numerical calculations by a finite-element analysis code. The deviation was less than 5~\% below mechanical resonance of the mass. We also checked if it is possible to compensate thermal motions caused by the same dissipation by changing the resonant condition of the light in the etalon. The compensation is possible but the noise level actually increases due to the circulation of the light in the etalon.

The Khalili Etalon has clear application, for example, in a future gravitational-wave detector. The method we developed can be used also to calculate thermal noise of partially transmissive optics like a beamsplitter. In the case one uses a separate-type Khalili cavity~\cite{Khalili}, instead of the etalon, our method is useful to calculate the influence of the back surface of the first mirror. 

There are several additional factors to be investigated before the Khalili Etalon is ready to be implemented in a future gravitational-wave detector. Several other kinds of thermal noise will be added by the use of the etalon. The etalon parameters should be optimized according to the magnitude of other thermal noise. From the experimental side, thermal lensing and light scattering might be a problem. These issues are out of scope of this paper but the investigation will be done in the near future~\cite{AlexeyKE}.

\section*{Acknowledgement}

We would like to thank Dr.~Kazuhiro Yamamoto for valuable comments. SH is supported by the Science and Technology Facilities Council (STFC). AG and SV are supported by LIGO team from Caltech and in part by NSF and Caltech grant PHY-0651036 and grant 08-02-00580 from Russian Foundation for Basic Research. DH and RN acknowledge the support of the German Science Foundation (DFG) under contract SFB Trans\-regio 7.

\begin{appendix}

\section{Mechanical transfer}
\label{app:mechtrans}
In this appendix we derive Eq. (\ref{eq:tot_noise}) with the parameter of mechanical transfer $\eta$. We start with the application of forces $\epsilon_1F_0$ to the front of the front coating and $\epsilon_2F_0$ to the back surface of the substrate ($\delta_1$ and $\delta_2$ in Fig. \ref{fig:mirror}). Due to the small spatial dimensions of the coatings we can assume stress and strain to be independent from the cylindrical axis in the coatings. This refers to a situation where the forces $\epsilon_1F_0$ and $\epsilon_2F_0$ are applied to the front and back surface of the substrate, respectively. In the approximation of thin coatings, the elastic response of the Khalili etalon is calculated in this paper by neglecting their influence. Then, a force acting on the front affects the back surface in exactly the same way as a force applied onto the back affects the front surface. Finally, the elastic energy in the coatings is obtained via transition conditions at the substrate coating boundary (see Eq. (\ref{eq:strain1}) and (\ref{eq:strain2})).

The linearity of the elastic equations allows us to divide the situation with two applied forces into two problems with only one force acting. This results in the following expressions $u_{i(j)}$ for the energy density at position $i$ due to force $j$. If only force $\epsilon_1F_0$ is acting on the front we obtain
\begin{eqnarray}
u_{1(1)}={\cal A}(\epsilon_1F_0)^2 \ , u_{2(1)}={\cal A}(\eta\epsilon_1F_0)^2 \ .
\end{eqnarray}
The same considerations for the force on the back $\epsilon_2F_0$ leads to
\begin{eqnarray}
u_{1(2)}={\cal A}(\eta\epsilon_2F_0)^2\ , u_{2(2)}={\cal A}(\zeta\epsilon_2F_0)^2 \ .
\end{eqnarray}
The coefficient $\eta$ represents the mechanical transfer of a force on one side of the etalon that increases/decreases the elastic energy of the coatings on the other side of the etalon. The symmetry of this problem is represented by the identical coefficient ${\cal A}$. But note the parameter $\zeta$ breaks the symmetry of the system. This parameter is introduced due to the fact that the stress component $T_{zz}$ is assumed to be zero in the back coating (i.e. the force $\epsilon_2F_0$ is acting on the boundary between substrate and coating) in contrast to the front coating.

Examining thermal noise of the etalon demands knowledge of the dissipated energy in the coating, which depends on thickness and mechanical loss $\phi$. We combine both effects into one averaged parameter $\overline{\phi}_i$ for each coating $i$. We arrive at the following expressions for the dissipated energies $\Delta U_i$,
\begin{eqnarray}
\Delta U_{1(1)}={\cal A}(\epsilon_1F_0)^2\overline{\phi}_1 \ , \Delta U_{2(1)}={\cal A}(\eta\epsilon_1F_0)^2\overline{\phi}_2 \ , \\
\Delta U_{1(2)}={\cal A}(\eta\epsilon_2F_0)^2\overline{\phi}_1 \ , \Delta U_{2(2)}={\cal A}(\zeta\epsilon_2F_0)^2\overline{\phi}_2 \ .
\end{eqnarray}

To combine these two results to the case of energy loss with two applied forces leads to an interferometric term. Stress and strain behave linearly with respect to the amplitude of the applied probes. But due to the spatial separation of the probe forces the associated energies will not add linearly. A close examination of the problem reveals an analogy to the model of the classical optical interferometry and finally yields
\begin{eqnarray}
\Delta U_1={\cal A}\left(\epsilon_1^2+\eta^2\epsilon_2^2+2\chi_1\eta\epsilon_1\epsilon_2 \right)F_0^2\overline{\phi}_1 \ ,
\label{equ:Ediss1}\\
\Delta U_2={\cal A}\left(\eta^2\epsilon_1^2+\zeta^2\epsilon_2^2+2\chi_2\eta\zeta\epsilon_1\epsilon_2 \right)F_0^2\overline{\phi}_2 \ .
\end{eqnarray}
In the above expressions $\chi_1$ and $\chi_2$ determine the strength of the superposition.
Due to the different conditions for $T_{zz}$ in the front and back coatings the calculation of energies differs leading to a different parameter $\chi$ for the total energy of the front and back coating.
Also note the relation $-1\leq\chi_1,\,\chi_2\leq1$ that limits the interval of possible energy values.
The exact values depend on the solution of the elastic equations for the Khalili etalon.

We want to compare this result to the situation of a conventional mirror. For this purpose we virtually remove the back coating ($\overline{\phi}_2=0$) and only consider the front coating of the etalon. This leads to the case of only a single force on the front surface ($\epsilon_1=1$ and $\epsilon_2=0$). Because our calculation only considers the response of the substrate, we can use Eq. (\ref{equ:Ediss1}) for this case, too. This yields the total dissipated energy
\begin{eqnarray}
\Delta U_{1,CM}={\cal A}F_0^2\overline{\phi}_1 \ .
\end{eqnarray} 
The same considerations hold for a conventional mirror only consisting of the back coating. With the help of Eq. (\ref{eq:FdT}) we can replace the dissipated energy by the noise spectrum and obtain Eq.~(\ref{eq:tot_noise}):
\begin{multline}
S_x(\Omega)=
\left(\epsilon_1^2+\eta^2 \epsilon_2^2+2\chi_1\,\epsilon_1\,\eta\epsilon_2\right)\tilde{S}_{1} \\
+\left(\eta^2\epsilon_1^2+\zeta^2\epsilon_2^2+2\chi_2\,\eta\epsilon_1\,\zeta\epsilon_2\right)\tilde{S}_{2} \ .\nonumber
\end{multline} 
$\tilde{S}_{1}$ and $\tilde{S}_{2}$ characterize the noise spectra of conventional mirrors possessing the etalon's front or back coating stack, respectively.

From Eq. (\ref{equ:Ediss1}) it is clear how to determine the coefficient $\eta$. Integrating the energy on the front coatings for two situations is sufficient: Calculate the dissipated energy of the front coating $\Delta U_1$ for a force applied (a) on the back surface ($\epsilon_1=0$, $\epsilon_2=1$) and (b) on the front surface ($\epsilon_1=1$, $\epsilon_2=0$). Dividing the result of (a) by the result of (b) gives the factor $\eta^2$. If we repeat this procedure for the energy dissipation in the back coating we obtain $\eta^2/\zeta^2$.

\end{appendix}

\bibliographystyle{junsrt}

\end{document}